 \documentclass[preprint2]{aastex}

\def\dsr{$\mathrm{R}_\odot$}

\shorttitle{Rotating Solar Models with Low Metal Abundances}
\shortauthors{Wuming Yang}

\begin{document}


\title{Rotating Solar Models with Low Metal Abundances as Good as Those with High Metal Abundances}
\author{Wuming Yang }
\affil{Department of Astronomy, Beijing Normal University, Beijing 100875, China}
\email{yangwuming@bnu.edu.cn, yangwuming@ynao.ac.cn}

\begin{abstract}
Standard solar models (SSM) constructed in accord with low metal abundances disagree
with the seismically inferred results. We constructed rotating solar models with low metal
abundances that included enhanced settling and convection overshoot.
In one of our rotating models, \textbf{AGSSr2a}, the convection overshoot allowed us
to recover the radius of the base of convection zone (CZ) at a level of $1\sigma$.
The rotational mixing almost completely counteracts the enhanced settling for
the surface helium abundance, but only partially for the surface heavy-element abundance.
At the level of $1\sigma$, the combination of rotation and enhanced settling
brings the surface helium abundance into agreement with the seismically inferred value
of $0.2485\pm0.0035$, and makes the model have better sound-speed and density profiles
than SSM constructed in accordance with high metal abundances. The radius of the base of
the CZ and the surface helium abundance of \textbf{AGSSr2a} are $0.713$ $R_{\odot}$ and $0.2472$, respectively;
the absolute values of the relative differences in sound speed and density
between it and the Sun are less than $0.0025$ and $0.015$,
respectively. Moreover, predicted neutrino fluxes of our model are comparable with
the predictions of previous research works.

\end{abstract}


\section{Introduction}

The previously accepted value of the ratio of heavy-element abundance to hydrogen
abundance of the Sun ($Z/X$), as derived from photospheric
spectroscopy is $0.0244$ \citep[hereafter GN93]{gre93} or
$0.0231$ \citep[hereafter GS98]{gre98}. Standard solar models (SSM)
constructed in accordance with that ratio are thought to reproduce the
seismically inferred results, such as the depth and helium abundance of the convection zone (CZ),
and sound-speed and density profiles. The seismically inferred radius of the base of the CZ (BCZ)
is $0.713\pm0.003$ \dsr{} \citep{chr91} or $0.713\pm0.001$ \dsr{} \citep{bas97},
the helium abundance in the solar envelope is $0.2485\pm0.0035$ \citep{bas04, ser10}, \textbf{
and the surface heavy-element abundance is in the range of $0.006-0.011$ \citep{voro14}
or in the range of $0.008-0.014$ \citep{buld17}.}

\cite{lod03} and \cite{asp04} analysed the solar photospheric abundances
and found that the value of $Z/X$ in the present-day photosphere is
in the range of about $0.0177-0.0165$, which is significantly lower than
was previously thought. SSMs constructed in accord with this low $Z/X$ disagree with
the earlier seismically inferred results \citep{bas04,
bah04b, bah05, mon04, mon06, tur04, guz05, cas07, yang07}.

\cite{asp09}, \cite{lod09}, and \cite{caf10} reanalyzed the solar photospheric abundances
and revised the heavy-element abundance to $Z=0.0134$ and $Z/X=0.0181$
\citep[hereafter AGSS09]{asp09}, $Z=0.0141$ and $Z/X=0.0191$ \citep{lod09},
and $Z=0.0154$ and $Z/X=0.0211$ \citep{caf10}. SSMs constructed in accordance with
these new low metal abundances also do not completely agree with the seismically
inferred results \citep{ser09, ser10, ser11, guz10, bi11, lep15} and the neutrino flux
constraints \citep{bah04a, tur04, tur10, tur11, tur12, lop13, lop14, yang16}.

Recently, \cite{von16} have claimed that the lower limit of solar metallicity ($Z_{\odot}$)
derived from in situ solar wind composition is $0.0196$, which is significantly larger
than those of AGSS09 and \cite{caf10}, but consistent with that of \cite{and89}.
This seems to aggravate the ``solar modeling problem" (or ``solar abundance problem").
\cite{ser16} and \cite{vag17} have shown that the abundances found by \cite{von16}
do not solve the problem but rather aggravate the surface helium abundance.
The ``solar modeling problem" thus remains unsolved.

\cite{asp04} suggested that increased diffusion and settling of helium and heavy
elements might be able to reconcile the low-Z models with helioseismology. It has
been found that enhanced settling can significantly improve sound-speed and density
profiles, but the surface helium abundance becomes too low and the position
of the BCZ too shallow \citep{bas04, mon04, guz05, yang07, yang16}.

The effects of rotation on solar models have been studied by many authors \citep{pin89,
yang06, yang07, tur10, yang16}, including the centrifugal effect
and rotational mixing. The former is insignificant for slowly rotating stars,
but even for slowly rotating stars, the latter plays an important role in shaping
their internal structures. Rotational mixing changes the distribution of
the mean molecular weight, so it can improve the sound-speed profile \citep{yang06}.

\cite{pro91} showed that macroscopic turbulent mixing can inhibit the settling of helium
by about $40\%$, but \cite{tho94} did not consider the effects of rotational
and turbulent mixing on the diffusion and settling of either helium or heavy elements
in their diffusion coefficients. Rotational mixing can partially counteract microscopic
diffusion and gravitational settling; so to obtain the same surface
helium abundance in a rotating solar model as in a non-rotating SSM,
the rates of microscopic diffusion and gravitational settling of \cite{tho94}
should be enhanced in the rotating model. 

In this work, our aim is to construct a solar model with low metal abundances
that is as good as the SSMs with high metal abundances, e.g. in good agreement
with the seismically inferred results. The paper is organized as follows.
The properties of different solar models are presented in Section 2, calculation
results are shown in Section 3, and the results are discussed and summarized in Section 4.

\section{Properties of Solar Models}

We used the Yale Rotation Evolution Code (YREC) \citep{pin89, yang07} in its rotation
and non-rotation configurations to compute solar models and the \cite{gue94}
pulsation code to calculate non-adiabatic oscillation frequencies of $p$-modes.
We used the OPAL EOS2005 tables \citep{rog02} and OPAL opacity tables \citep{igl96} with
GN93, GS98, and AGSS09 mixtures, but supplemented by the \cite{fer05} opacity
tables at low temperature as reconstructed with the GN93, GS98, and AGSS09 mixtures.
We computed the diffusion and settling of both helium and heavy elements by using
the formulas of \cite{tho94}; and we calculated the nuclear reaction rates by
using the subroutine of \cite{bah92} and \cite{bah95, bah01}, but updated by the reaction
data in \cite{ade98}, \cite{gru98}, and \cite{mar00}. Convection was determined by 
the Schwarzschild criterion and treated according to the standard mixing-length theory.
The depth of convection overshoot region was determined by $\delta_{\rm ov}H_{p}$,
where $\delta_{\rm ov}$ is a free parameter and $H_{p}$ is the local pressure
scale height. Throughout much of the overshoot region, motion is no doubt
sufficiently rapid to maintain the temperature gradient essentially at the adiabatic
value \citep{chr91}. Thus, the overshoot region is assumed to be both fully mixed
and adiabatically stratified.

All models are calibrated to the present solar luminosity $3.844\times10^{33}$
erg $\mathrm{s}^{-1}$, radius $6.9598\times10^{10}$ cm, mass $1.9891\times10^{33}$ g,
and age $4.57$ Gyr. The initial hydrogen abundance $X_{0}$, metallicity $Z_{0}$,
and mixing-length parameter $\alpha$ are free parameters adjusted to match
the constraints of luminosity and radius within around $10^{-5}$ \textbf{and an
observd $(Z/X)_{s}$.} The initial rotation rate ($\Omega_{i}$) of our rotating models 
is also a free parameter but adjusted to obtain a surface equatorial velocity of 
about $2.0$ km s$^{-1}$ at the age of 4.57 Gyr. The distribution of the initial 
angular velocity is assumed to be uniform, and angular-momentum loss from the CZ 
due to magnetic braking is computed by using Kawaler's relation \citep{kaw88, cha95}.
The values of the parameters are summarized in Table \ref{tab1}.

In the rotating models, the redistributions of angular momentum and chemical compositions
are treated as a diffusion process, i.e.
    \begin{equation}
       \frac{\partial \Omega}{\partial t}=f_{\Omega}
       \frac{1}{\rho r^{4}}\frac{\partial}{\partial r}(\rho r^{4}D
       \frac{\partial \Omega}{\partial r}) \,
      \label{diffu1}
    \end{equation}
for the transport of angular momentum and
    \begin{equation}
    \begin{array}{lll}
        \frac{\partial X_{i}}{\partial t}&=&f_{c}f_{\Omega}\frac{1}{\rho r^{2}}
       \frac{\partial}{\partial r}(\rho r^{2}D\frac{\partial X_{i}}
        {\partial r})\\
        & &+(\frac{\partial X_{i}}{\partial t})_{\rm nuc}-\frac{1}
       {\rho r^{2}}\frac{\partial}{\partial r}(f_{0}\rho r^{2}X_{i}V_{i}) \,
    \end{array}
      \label{diffu2}
    \end{equation}
for the change in the mass fraction $X_{i}$ of chemical species $i$,
where $D$ is the diffusion coefficient caused by rotational instabilities.
Those instabilities include the Eddington circulation, the Goldreich-Schubert-Fricke
instability \citep{pin89}, and the secular shear instability \citep{zahn93}.
In our calculations, the values of $f_{\Omega}$ and $f_{c}$ are $1$ and $0.03$, respectively,
though the value of $f_{c}$ is commomly put at about $0.02-0.05$ \citep{cha92, yang06, bro11}
to explain solar lithium depletion \citep{pin89} and the extended
main-sequence turnoffs in the color-magnitude diagram of intermediate-age massive
star clusters in the Large Magellanic Cloud \citep{yang13}.
The inhibiting effect of chemical gradients on the efficiency of rotational mixing
is taken into consideration in our models and is regulated by the parameter $f_{\mu}$
with a value of $0.1$ as described in \cite{pin89}.
The second term on the \textbf{right-hand side} of Equation (\ref{diffu2}) represents the change
caused by the nuclear reaction. In that equation, $V_{i}$ is the
velocity of microscopic diffusion and settling given by \cite{tho94}; and $\rho$
is density. The parameter $f_{0}$ is a constant. In standard cases,
the value of $f_{0}$ is equal to $1$; but in an enhanced settling model,
it is larger than $1$. Moreover, in our rotating models, a convection overshoot
of $\delta_{ov}=0.1$ is required to reproduce the seismically inferred depth of the CZ.

\section{Calculation Results}
\subsection{Solar models with high metal abundances}
\subsubsection{Standard solar models with high metal abundances}

We first constructed two SSMs with high metal abundances --- GN93M and GS98M ---
by using GN93 mixture opacities and GS98 mixture opacities, respectively. The radius $r_{cz}$
of BCZ, the surface helium abundance $Y_{s}$, and other fundamental parameters of
the models are shown in Table \ref{tab1}; and predicted and detected solar neutrino
fluxes in Table \ref{tab2}. We compared the neutrino fluxes computed
from our models with those calculated from the models BP04 \citep{bah04a} and
SSeM \citep{cou03}.

The sound speed and density of our models were compared with those inferred by
\cite{bas09} using the data from the Birmingham Solar-Oscillations Network
\citep{cha96} and the Michelson Doppler Imager \citep{sch98} (see Figure \ref{fig1}). 
The absolute values of relative sound-speed difference, $\delta c/c$, and density difference,
$\delta\rho/\rho$, between the two SSMs and the Sun are less than $0.0041$ and $0.027$, 
respectively.

Figure \ref{fig2} shows that the ratios of small to large separations, $r_{02}$
and $r_{13}$, of our SSMs agree with those calculated from observed frequencies
of \cite{cha99b} or \cite{gar11}. The neutrino fluxes computed from GS98M are
comparable with those calculated from BP04 \citep{bah04a} and those detected
by \cite{ahm04} and \cite{bel11, bel12} (see Table \ref{tab2}). Therefore, we chose
GS98M as our best SSM with high metal abundances and went \textbf{on} to construct solar 
models with low metal abundances and compare them with this high abundance model.

\subsubsection{Enhanced diffusion and rotating models with high metal abundances}

\begin{figure}
\includegraphics[angle=-90, scale=0.6]{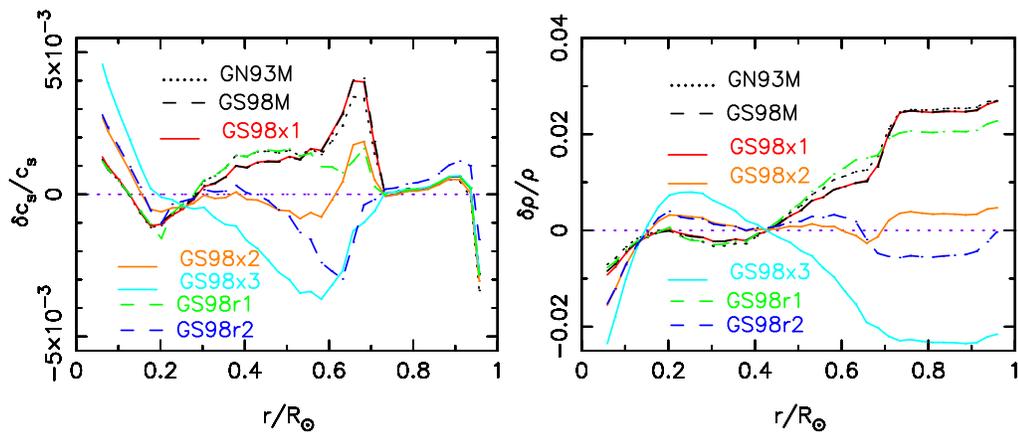}
\caption{The relative sound-speed and density differences, in the sense of the
(Sun-Model)/Model, between solar models and seismically inferred results.
The seismically inferred sound speed and density are given in \cite{bas09}.
\textbf{The red line coincides with the black dashed line.}
\label{fig1}}
\end{figure}

\begin{figure}
\includegraphics[angle=-90, scale=0.6]{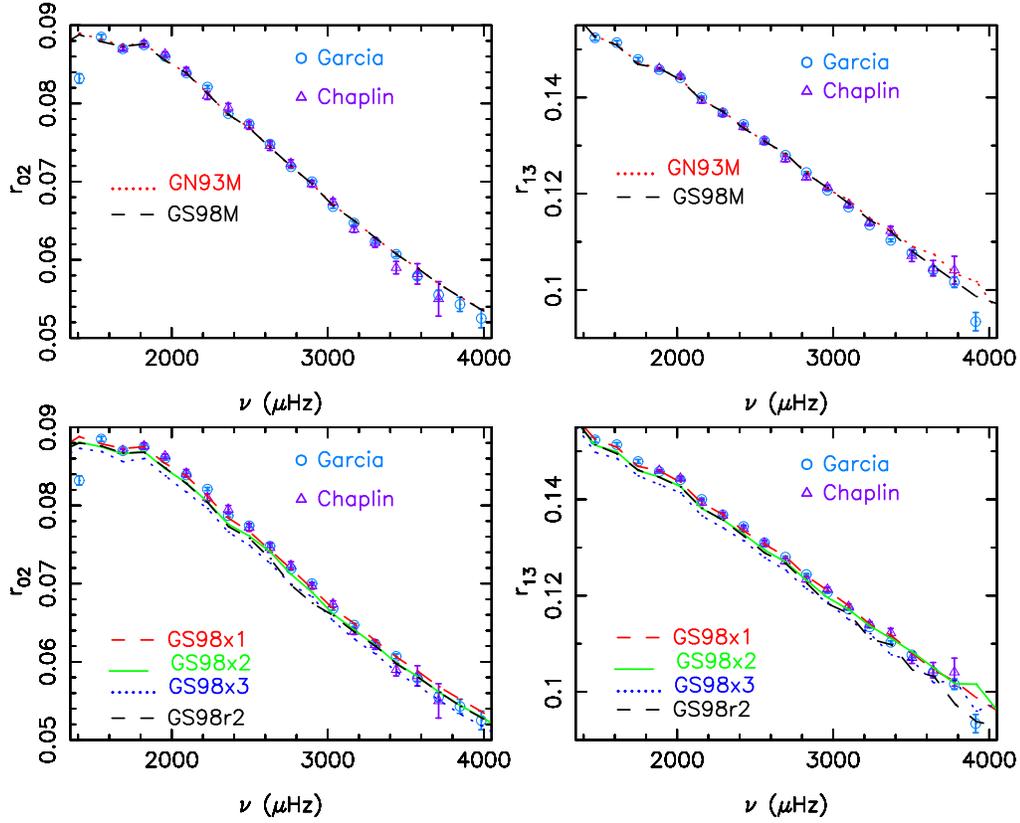}
\caption{The distributions of the ratios of small to large separations,
$r_{02}$ and $r_{13}$, as a function of frequency. The circles and triangles
show the ratios calculated from the frequencies observed by GOLF \& VIRGO \citep{gar11}
and BiSON \citep{cha99b}, respectively.
\label{fig2}}
\end{figure}

We first needed to study the counteraction between rotational mixing and the diffusion
of elements; so we constructed three enhanced settling models GS98x1, GS98x2, and GS98x3
and two rotating models GS98r1 and GS98r2. We enhanced the rates of element diffusion
and settling in the first three models and in GS98r2 by applying a straight multiplier
($f_{0}$) to the diffusion velocity of \cite{tho94}, as \cite{bas04}, \cite{mon04}, \cite{guz05},
and \cite{yang07} \textbf{have done,} despite the fact that there is no obvious physical justification
for such multiplier \citep{bas04, guz05}. For GS98x1, we increased the rates of
element diffusion and settling by $10\%$, which did not affect its properties
except for the surface helium abundance (see Figure \ref{fig1} and Table \ref{tab1}).
For GS98x2, we enhanced the rates by $50\%$. This significantly improved its sound-speed
and density profiles (see Figure \ref{fig1}), but led to the fact that its surface helium
abundance is too low. Moreover, the multiplier is too large. The theoretical error
of the gravitational settling rate is of the order of about $15\%$ \citep{tho94}
in cases that rotational and turbulent mixing is neglected. For GS98x3, we increased
the rates by $100\%$; but we found that the surface helium abundance, \textbf{radius $r_{cz}$,} 
sound-speed and density profiles of GS98x3 become worse than those of GS98M and GS98x2. 
This implies that it is unlikely that the rates can be increased by $100\%$.

The effects of rotation can improve sound-speed and density profiles \citep{yang06}.
The profiles of rotating model GS98r1 is better than those of our best SSM with
high metal abundance, GS98M. However, rotational mixing raised the surface helium
abundance of GS98r1 to $0.2548$, which is higher than the seismically inferred value of
$0.2485\pm0.0035$. In order to reconcile the high surface helium abundance 
with that inferred, we needed to enhance the rates of element diffusion 
and settling in the rotating model.

We enhanced the rates of element diffusion and settling of rotating model GS98r2
by $50\%$ as high as that of enhanced settling model GS98x2 mentioned above.
The absolute values of the relative sound-speed and density differences between GS98r2
and the Sun were found to be less than $\mathbf{0.003}$ and $\mathbf{0.0155}$, 
respectively, which is better than those of GS98M (see Figure \ref{fig1}).
The predicted neutrino fluxes of GS98r2 are almost the same as those of GS98M;
but the surface helium abundance of $\mathbf{0.2466}$ for GS98r2 is larger than
the $0.2315$ of GS98x2, though smaller than the $0.2548$ of GS98r1.

\begin{figure}
\includegraphics[angle=-90, scale=0.6]{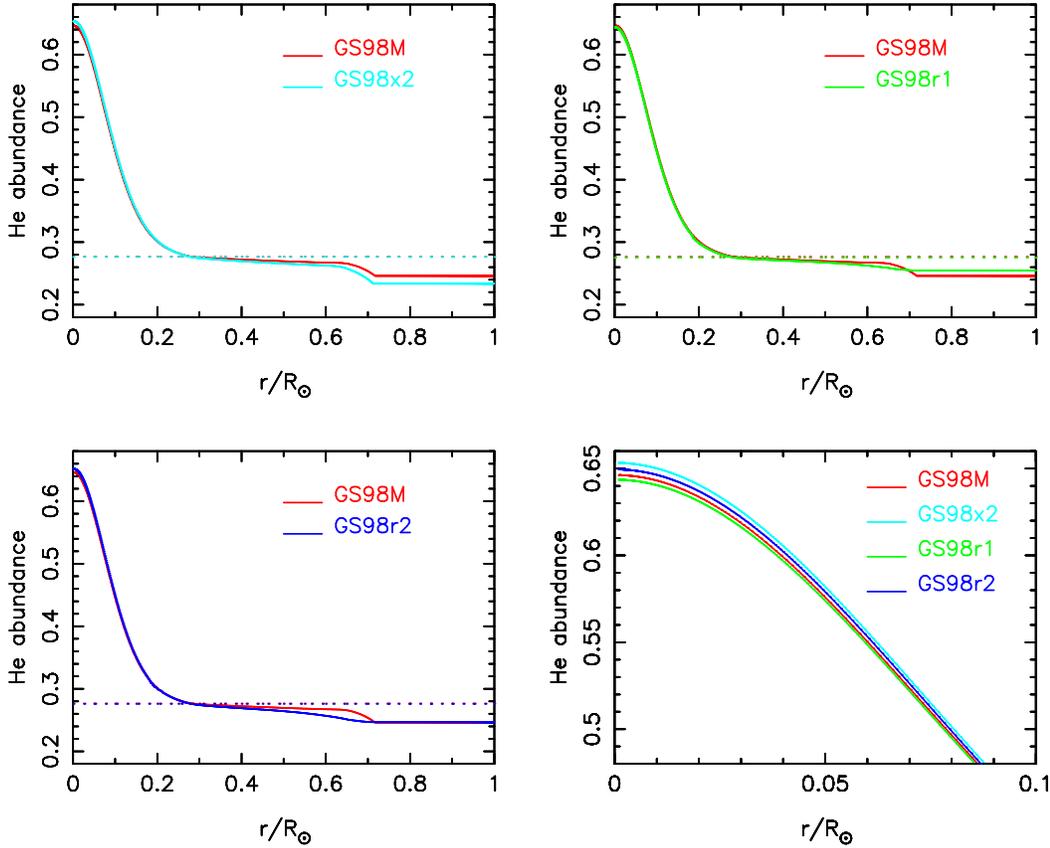}
\caption{Distributions of He abundance of models as a function of radius.
The dotted lines show the initial heliun abundances of the models. The central
He abundances of GS98x2 and GS98r2 are higher than that of GS98M, which is caused
by the enhanced settling. The initial helium abundances of GS98M, GS98x2, GS98r1,
and GS98r2 are $0.27669$, $\mathbf{0.27635}$, $0.27518$, and $\mathbf{0.27590}$, respectively. 
\label{fig3}}
\end{figure}

The initial helium abundances of GS98M, GS98x2, GS98r1, and GS98r2 were $0.27669$,
$\mathbf{0.27635}$, $0.27518$, and $\mathbf{0.27590}$, respectively. 
Gravitational settling reduced the surface helium abundance of GS98M 
by about $11.7\%$ below its initial value, and that of GS98x2, GS98r1, 
and GS98r2 by around $\mathbf{15.5}\%$, $7.4\%$, and $\mathbf{10.6\%}$, respectively.
Figure \ref{fig3} shows the distributions of helium abundances of the models as a function
of radius. Due to the enhanced settling, the central helium abundance of GS98x2
was higher than that of GS98M, but its surface helium abundance lower. Rotational mixing
made the surface helium abundance of GS98r1 higher than that of GS98M.
In GS98r2, the effect of rotational mixing on the surface helium abundance was almost
counteracted by the increased settling. Thus its surface helium abundance is almost consistent
with that of GS98M. This implies that the velocity of element diffusion and settling
should be increased by roughly $50\%$ to obtain the same surface
helium abundance in a rotating solar model as in a non-rotating SSM.

\subsection{Solar models with low metal abundances}

\subsubsection{Standard and enhanced diffusion models}
\begin{figure}
\includegraphics[angle=-90, scale=0.6]{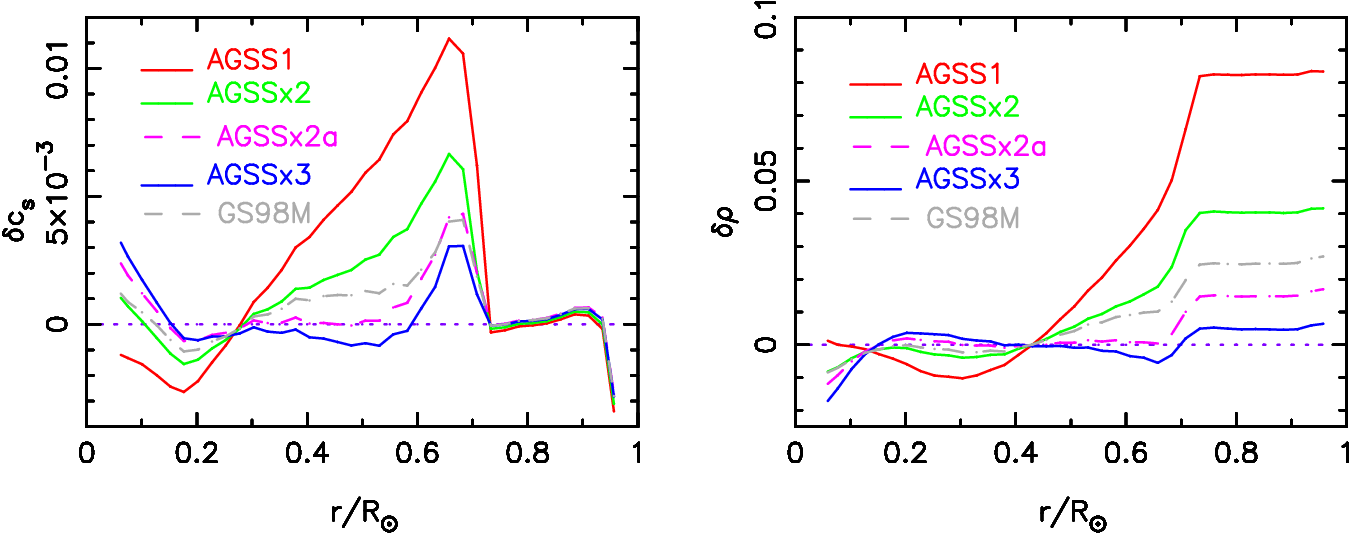}
\caption{The relative sound-speed and density differences, in the sense of the
(Sun-Model)/Model, between solar models and seismically inferred results.
The seismically inferred sound speed and density are given in \cite{bas09}.
\label{fig4}}
\end{figure}

The SSM AGSS1 constructed with AGSS09 mixture opacities disagrees with the seismically
inferred results, as has generally been found. In order to study the effects of enhanced settling,
we constructed two models with low metal abundances --- AGSSx2 and AGSSx3 --- with different
multipliers (see Table \ref{tab1}). \textbf{Models AGSSx2 and AGSSx3 had the Z/X determined 
by \cite{lod09}. The enhanced settling required a high $Z_{0}$ to reproduce the observed Z/X.} 
The sound-speed and density profiles of AGSSx2 and AGSSx3 are obviously better than those 
of \textbf{AGSS1} (see Figure \ref{fig4}). This implies that enhanced settling aids in improving 
solar models. \textbf{We also constructed a model AGSSx2a with the surface heavy-element abundance 
determined by \cite{caf10}. The sound-speed and density profiles of AGSSx2a are better than those 
of AGSSx2, even better than those of GS98M (see Figure \ref{fig4}). This can be partially attributed to 
the increased settling and the initial metal abundance.} The value of $Z_{0}$ of \textbf{AGSSx2a}
is larger than \textbf{that of AGSSx2.} The inferred sound speed \textbf{and density are} 
more easily reproduced by a model with a high initial metal abundance. The enhanced settling results 
in the main differences between \textbf{AGSS1 and AGSSx2. But the higher $Z_{0}$ of AGSSx2a leads}
to the differences between \textbf{AGSSx2 and AGSSx2a}. 

\begin{figure}
\includegraphics[angle=-90, scale=0.6]{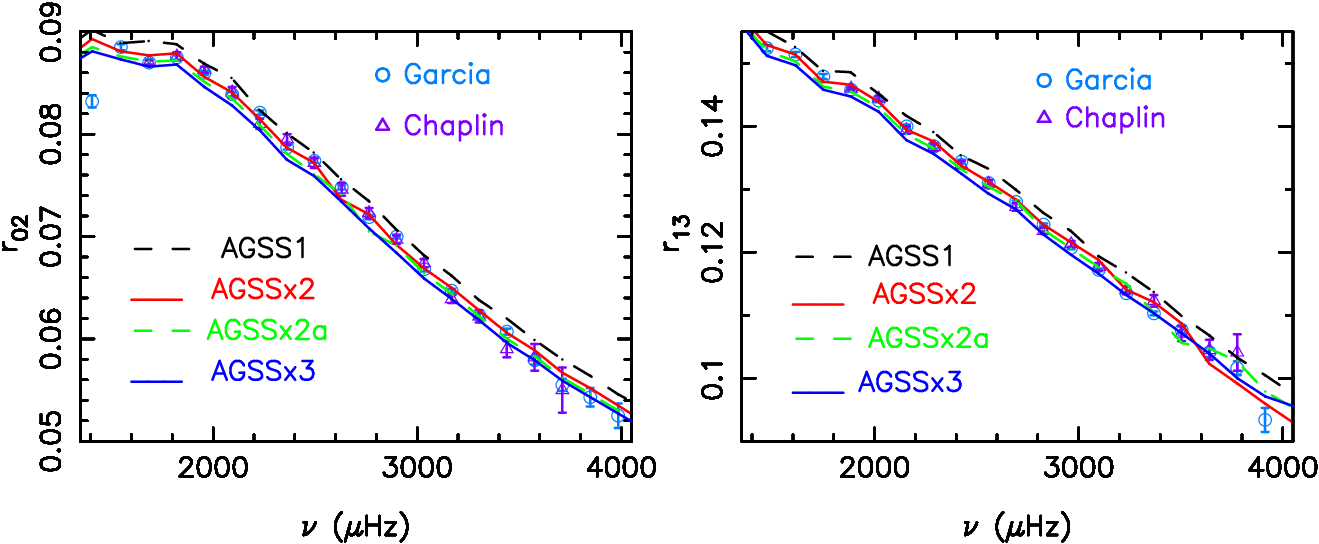}
\caption{The distributions of the ratios of small to large separations,
$r_{02}$ and $r_{13}$, as a function of frequency. The circles and triangles
show the ratios calculated from the frequencies observed by GOLF \& VIRGO \citep{gar11}
and BiSON \citep{cha99b}, respectively.
\label{fig5}}
\end{figure}

However, Table \ref{tab1} shows that the larger the value of $f_{0}$, the lower the surface 
helium abundance. The surface helium abundances of AGSSx2, \textbf{AGSSx2a,} and AGSSx3
are lower than the seismically inferred value. Moreover, Figures \ref{fig2} and \ref{fig5} 
show that the $r_{02}$ and $r_{13}$ of the enhanced settling models deviate from
the observed ones with an increase in the multiplier. Only the $r_{02}$ and $r_{13}$ of 
AGSSx2 and \textbf{AGSSx2a} whose diffusion velocities are enhanced by $50\%$ agree with
the observed ones. This indicates that, as has been found by \cite{bas04}, \cite{guz05},
and \cite{yang16}, enhanced settling alone cannot solve the solar modeling problem.

\subsubsection{Rotating models as good as GS98M}
Considering that rotational mixing can counteract enhanced settling to a certain degree,
we constructed \textbf{three} rotating models --- AGSSr2, \textbf{AGSSr2a,} and AGSSr3 --- 
with AGSS09 mixture opacities. \textbf{Models AGSSr2 and AGSSr3 had the surface heavy-element 
abundance derived by \cite{lod09}, but AGSSr2a had that determined by \cite{caf10}.} 
Table \ref{tab1} lists their fundamental parameters. Figure \ref{fig6} 
shows that AGSSr3 has a better density profile, but it cannot reproduce the observed $r_{02}$ 
and $r_{13}$ (see Figure \ref{fig7}) and the seismically inferred surface helium abundance
(see Table \ref{tab1}). This is due to the fact that the enhanced settling is too large to 
be counteracted by rotational mixing. Thus AGSSr3 is not a good solar model.

\textbf{ 
The radius $r_{cz}$, sound-speed and density profiles of AGSSr2 are as good as
those of GS98M (see Table \ref{tab1} and Figure \ref{fig6}). It also reproduces
the observed $r_{02}$ and $r_{13}$ (see Figure \ref{fig7}). Moreover, the neutrino
fluxes computed from AGSSr2 are more consistent with the detected ones than other 
models (see Table \ref{tab2}). However, its surface helium abundance of $0.2425$ 
is slightly lower than the seismically inferred value of $0.2485\pm0.035$.
}

Table \ref{tab1} shows that \textbf{AGSSr2a} reproduces the seismically inferred surface
helium abundance and the radius $r_{cz}$ at the level of $1\sigma$. It also reproduces
the observed $r_{02}$ and $r_{13}$ (see Figure \ref{fig7}).
Furthermore, \textbf{AGSSr2a} has better sound-speed and density profiles
than GS98M (see Figure \ref{fig6}). The absolute values of
$\delta c/c$ and $\delta\rho/\rho$ between \textbf{AGSSr2a} and the Sun are less
than $0.0025$ and $0.015$, respectively. Predicted neutrino fluxes of \textbf{AGSSr2a}
are comparable with the predictions of BP04 and GS98M and the detected values (see Table \ref{tab2}).
The surface metallicity of $Z=0.0152$ is higher than that determined
by AGSS09, but is in good agreement with the value of $0.0154$ estimated by \cite{caf10}.
\textbf{AGSSr2a} is thus a solar model as good as, and to a certain degree even better than,
our best model with high metal abundance, GS98M.
It has better sound-speed and density profiles, and its surface
helium abundance of $0.2472$ is in good agreement with $0.2485\pm0.0035$.
Moreover, the radius of the BCZ of \textbf{AGSSr2a} --- $0.713$ \dsr{} --- is also
consistent with the seismically inferred value of $0.713\pm0.001$ \dsr{} \citep{bas97}.
Therefore, we chose \textbf{AGSSr2a} as our overall best model. This rotating solar model with
low metal abundances is comparable to, or even slightly better than the SSMs with high metal abundances.

Gravitational settling reduces the surface helium abundance by about $11.7\%$
in GS98M and around $10.8\%$ in \textbf{AGSSr2a} below their initial
values, which shows that for the surface helium abundance the enhanced
settling was counteracted by rotational mixing. Therefore, at the level
of $1\sigma$ the surface helium abundance of \textbf{AGSSr2a} is in agreement
with the seismically inferred value. Moreover, the gravitational settling reduces
the heavy-element abundance in the CZ by around $10.1\%$ in GS98M,
$14.5\%$ in \textbf{AGSSx2a}, and $13.8\%$ in \textbf{AGSSr2a} below their
initial values. For heavy elements, the enhanced settling is only partially
counteracted by rotational mixing, which, as Equation (\ref{diffu2}) shows,
is dependent on the radial gradient of elements. Due to the fact that central 
hydrogen burning is continuously producing helium, the gradient of helium abundance
in the Sun arises from gravitational settling along with central hydrogen burning.
However, the gradient of heavy elements results only from gravitational settling. 
Even in the radiative region with $r \gtrsim 0.2$ \dsr{}, the gradient of helium 
abundance is markedly larger than that of the heavy elements (see Figure \ref{fig8}), 
which leads to the fact that the effect of rotational mixing
on helium abundance is larger than that on heavy-element abundance. In other words,
rotational mixing can more efficiently inhibit the settling of helium than
of heavy elements. As a consequence, in the rotating model --- \textbf{AGSSr2a}, the effect of
the enhanced settling on the surface helium abundance is almost counteracted
by rotational mixing, but that effect on the heavy-element abundance in the CZ
is only partially offset by the mixing. This leads to the fact that the surface helium
abundance of \textbf{AGSSr2a} is almost the same as that of GS98M, but the heavy-element abundance
in the CZ of \textbf{AGSSr2a} is lower.

\subsubsection{Internal rotations of models}

Figure \ref{fig8} shows the distributions of the diffusion coefficient $D$ of
rotational instabilities, angular velocity, helium abundance, and metal abundance of model 
\textbf{AGSSr2a} at different ages as a function of radius. The large gradient of angular velocity
at around $r=0.2$ \dsr{} leads to an increase in $D$ (see panels $a$ and $b$ of 
Figure \ref{fig8}). The diffusion coefficient $D$ is dependent on evolutionary states.
The central angular velocity is about $5$ times as large as the surface angular velocity.
Seismically inferred results show that the solar angular velocity profile
is flat in the radiative region of the Sun \citep{cha99a}. The internal
rotation rate of \textbf{AGSSr2a} is higher than that of the Sun and its total
angular momentum is about $3.97\times10^{48}$ g cm$^{2}$ s$^{-1}$,
which is larger than that of the Sun as inferred by \cite{kom03}:
 $(1.94\pm0.05)\times10^{48}$ g cm$^{2}$ s$^{-1}$. A decrease in the
initial angular velocity cannot bring the total angular momentum into
agreement with the seismically inferred one. For example, the initial angular velocity
of GS98r1 is lower than that of \textbf{AGSSr2a}; and the total angular
momentum of GS98r1 is $3.86\times10^{48}$ g cm$^{2}$ s$^{-1}$.

The instabilities considered in this work mainly take effect in the region with
$r \gtrsim 0.2$ \dsr{}. The core contracts as the Sun evolves and so
rotates faster and faster. These lead to too high internal angular
velocities in the rotating models and a total angular momentum that
is larger than the seismically inferred one. This indicates that there 
must be a mechanism of angular momentum transport affecting the inner 
layers of the models' radiative region. But it is missing.
A gravity wave \citep{cha05, gar07} or other mechanism \citep{den08} might be
a candidate for the angular momentum transport in these layers. \textbf{Such 
additional angular-momentum transport would obviously change the evolution 
of the internal rotation rate and hence the effects of turbulent diffusion 
on the composition.}

\section{Discussion and Summary}

A convection overshoot is required in rotating models to match the inferred
depth of the CZ; but it is not required in the increased settling models, 
\textbf{AGSSx2a} and AGSSx3. This is partially due to the initial element abundances 
and rotational mixing, which lead to rotating models having different distributions
of elements. \textbf{An increased settling requires a higher $Z_{0}$.} 
The seismically inferred sound speed and depth of the CZ are more easily
reproduced by a model with a higher metallicity. The enhanced settling model, 
\textbf{AGSSx2a}, has a higher $Z_{0}$. \textbf{For AGSSx2,} a convection overshoot 
of $\delta_{\rm{ov}}=0.1$ is required to match the inferred depth of the CZ.

In our models, rotational mixing plays a major role in recovering the surface
helium abundance and sound-speed profile. The mixing is dependent on
the diffusion coefficient $D$ of rotational instabilities, including those 
described by \cite{pin89}, and the secular shear instability of \cite{zahn93}.
The occurrence of rotational instabilities is thought to be related to the critical 
Reynolds number $R_{crit}$, the value of which in our calculations is $1000$, in line
with the default of \cite{end78} and \cite{pin89}. The diffusion
coefficient $D$ is dependent on the evolutionary states. Different scenarios
for the evolution of rotation could lead to different turbulent diffusion \textbf{
to a certain degree, and hence also the turbulent diffusion and composition profiles.}

According to the works of \cite{cha92} and \cite{mae98}, angular momentum
is transported through advection (by meridional circulation velocity $\mathbf{u}$) and
through turbulent diffusion; but the vertical transport of chemical elements can still
be described by a diffusion equation. The advection term in the equation of angular momentum
transport of \cite{mae98} is not considered in the YREC. The effect of Eddington circulation
on the redistribution of angular momentum is treated as a diffusion process in the YREC.
This could affect the angular velocity profiles of our models to a certain degree.

\cite{yang07}, \cite{tur10}, and \cite{yang16} studied the effects of rotation on solar models
and found that the effects of rotation alone cannot solve the solar modeling problem.
A combination of rotation and increased settling is required in our models to explain
both the low-metal abundance and the seismically inferred results. The values of multiplier 
($f_{0}$) in \cite{yang07} and \cite{yang16} are greater than or equal to $2$. In that case, 
rotational mixing cannot counteract the increased settling, not even for the surface helium abundance.
Different from the earlier models of \cite{yang07} and \cite{yang16}, in \textbf{AGSSr2a}, in which
the multiplier is $1.5$, rotational mixing almost completely counteracts the effect of 
the increased settling for its surface helium abundance. The sound-speed profile, surface helium abundance, 
depth of the CZ, and $r_{02}$ and $r_{13}$ of \textbf{AGSSr2a} are thus in better agreement with seismically 
inferred ones than those of the earlier models, and the neutrino fluxes computed from \textbf{AGSSr2a}
are more consistent with those detected by \cite{ahm04} and \cite{bel11, bel12} and predictions 
of BP04 \citep{bah04a} than those calculated from the earlier models. 
Therefore, \textbf{AGSSr2a} is significantly better than the earlier models.

However, there is no obvious physical justification for the multiplier.
Rotational mixing can partially counteract the microscopic diffusion and gravitational 
settling. Therefore, in order to obtain the same surface helium abundance
in a rotating solar model as in a non-rotating SSM, the rates of microscopic diffusion
and gravitational settling need to be enhanced in the rotating model.

Convection overshoot also can lead to an increase in the surface helium abundance
\citep{mon06, cas07}. But comparing with the effect of rotational mixing, \cite{yang16}
shows that the effect of a convection overshoot of $\delta_{ov}=0.1$ on the surface helium
abundance is negligible. The main effect of the overshoot is to bring the depth of the CZ 
into agreement with the seismically inferred one. If the overshoot is not considered 
in rotating models, the depth of the CZ of the models will be too shallow.

In our calculations, we computed the nuclear reaction rates by using the subroutine
of \cite{bah92}, but updated by \cite{bah95} and \cite{bah01} using new reaction data.
The discrepancy between the $hep$ neutrino flux of our models and that of BP04
can be attributed to the nuclear cross section factor $S_{0}(hep)$ \citep{bah01}.
We used the factor of \cite{mar00} for $hep$ neutrino flux,
which affects only the flux of $hep$ neutrinos, not those of
other neutrinos \citep{yang16}. The factors of nuclear reaction cross sections 
could affect the predicted neutrino fluxes and lead to the fact
that the neutrino fluxes of our models are more consistent with those of
BP04 \citep{bah04a} rather than those of SSeM \citep{tur11a}. The $^{7}$Be
neutrino fluxes predicted by \textbf{GS98M and AGSSr2a}
agree with the detected one and the prediction
of BP04 at the level of $2\sigma$; and the fluxes of $pp$, $pep$, $^{8}$B,
$^{13}$N, $^{15}$O, and $^{17}$F neutrinos computed from \textbf{AGSSr2a} are in good
agreement with the predictions of BP04.

\begin{figure}
\includegraphics[angle=-90, scale=0.7]{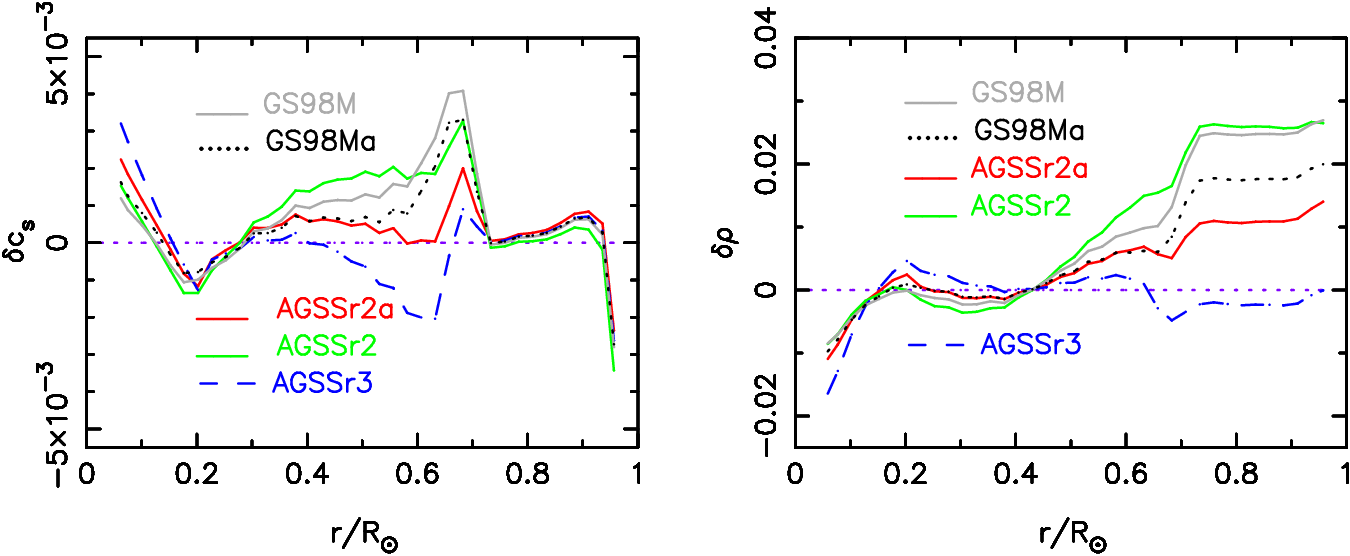}
\caption{The relative sound-speed and density differences, in the sense of the
(Sun-Model)/Model, between solar models and seismically inferred results.
The seismically inferred sound speed and density are given in \cite{bas09}.
\label{fig6}}
\end{figure}

\begin{figure}
\includegraphics[angle=-90, scale=0.7]{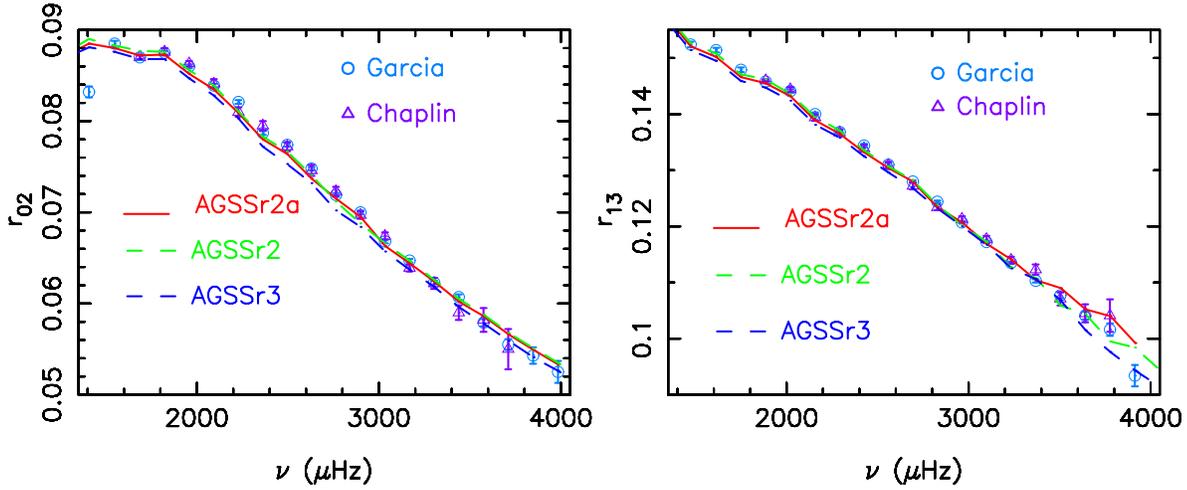}
\caption{The distributions of the ratios of small to large separations,
$r_{02}$ and $r_{13}$, as a function of frequency. The circles and triangles
show the ratios calculated from the frequencies observed by GOLF \& VIRGO \citep{gar11}
and BiSON \citep{cha99b}, respectively.
\label{fig7}}
\end{figure}

\begin{figure}
\includegraphics[angle=0, scale=0.23]{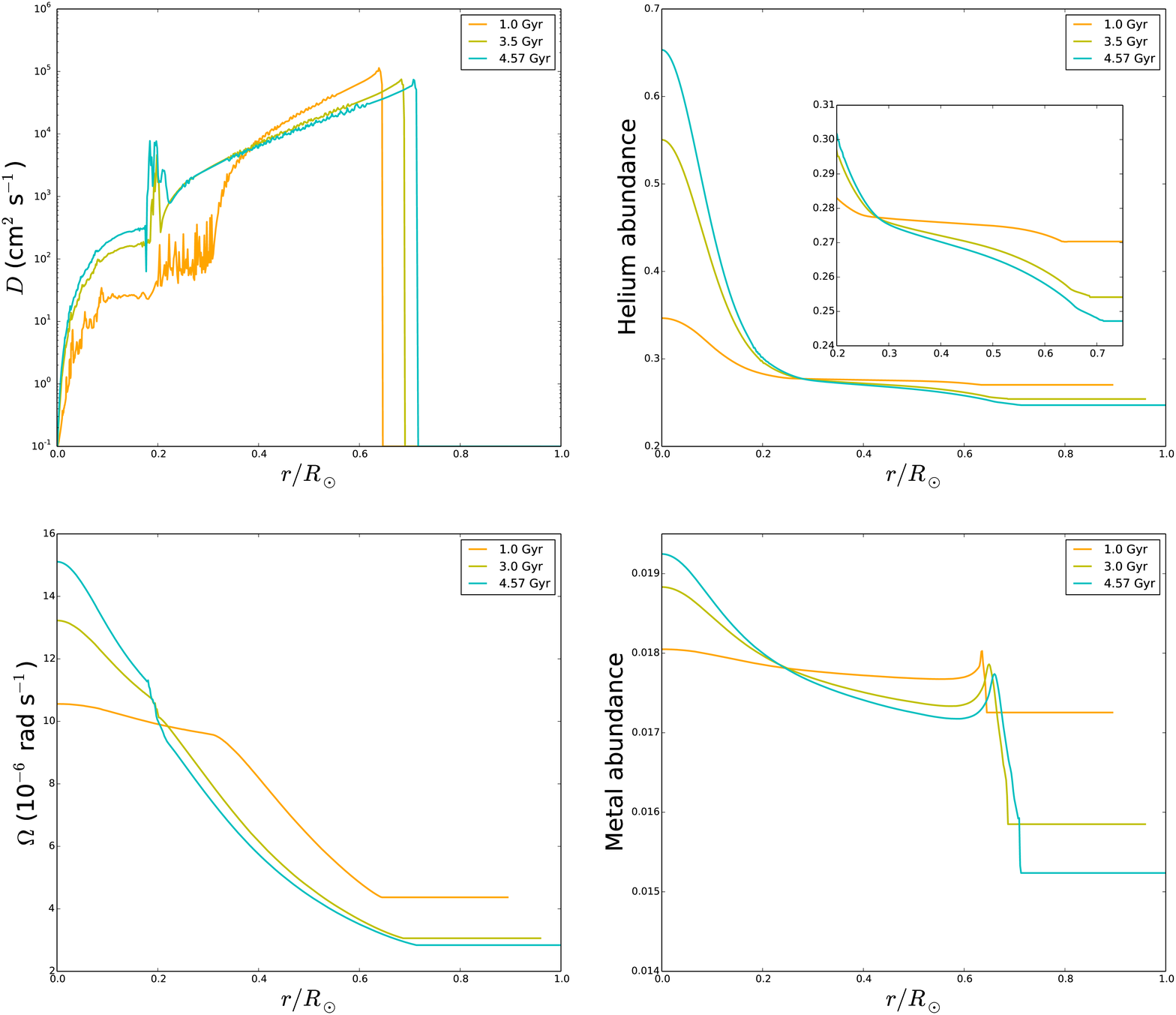}
\caption{Distributions of diffusion coefficient $D$, angular velocity,
helium abundance, and metal abundance of model \textbf{AGSSr2a} at different ages as a function of radius.
The different color lines refer to different ages.
\label{fig8}}
\end{figure}

If the surface metallicity of SSM with GS98 mixtures is increased to $0.01809$,
we can obtain model GS98Ma with $r_{cz}=0.715$ \dsr{} and $Y_{s}=0.2483$,
and with absolute values of $\delta c/c$ and $\delta\rho/\rho$ that are less than $0.0033$
and $0.020$, respectively. The neutrino fluxes predicted by
GS98Ma are almost the same as the predictions of \textbf{AGSSr2a}, but \textbf{AGSSr2a} has
better sound-speed and density profiles than GS98Ma \textbf{(see Figure \ref{fig6})}.
Thus \textbf{AGSSr2a} is better than GS98Ma and is still the best model in our calculations.

In this work, we constructed rotating solar models with the AGSS09 mixture opacities
where the effects of enhanced settling and convection overshoot were included.
We obtained a rotating model \textbf{AGSSr2a} that is better than SSM GS98M and
the earlier rotating models of \cite{yang07} and \cite{yang16}. The surface helium
abundance of \textbf{AGSSr2a} is $0.2472$, which agrees with the seismically
inferred value at the level of $1\sigma$. The radius of the BCZ of
$0.713$ \dsr{} is also consistent with the seismically
inferred value of $0.713\pm0.001$ \dsr{} \citep{bas97}. The absolute values
of $\delta c/c$ and $\delta\rho/\rho$ are less than $0.0025$ and $0.015$
for \textbf{AGSSr2a}, respectively. The sound speed
and density of \textbf{AGSSr2a} are more consistent with the seismically
inferred values \citep{bas09} than those of GS98M. The surface metal abundance
of \textbf{AGSSr2a} is $0.0152$, which is higher than the value evaluated by AGSS09,
but in agreement with the $0.0154$ determined by \cite{caf10}. The initial helium abundance
of \textbf{AGSSr2a} is $0.27721$, which is consistent with the
value of $0.273\pm0.006$ inferred by \cite{ser10}.

The main effect of convection overshoot is to bring the radius of the BCZ
into agreement with the seismically inferred value at the level of $1\sigma$.
In the model, the velocities of diffusion and settling of
helium and heavy elements of \cite{tho94} are enhanced by $50\%$.
Rotational mixing almost completely counteracts the increased settling for 
the surface helium abundance, but only partially for the heavy-element abundance
in the CZ. As a consequence, the surface helium abundance
of \textbf{AGSSr2a} is almost as high as that of GS98M, but the surface heavy-element
abundance is lower. A combination of the enhanced settling
and the effects of rotation allows us to recover the surface helium
abundance at the level of $1\sigma$ and makes the model have better
sound-speed and density profiles than GS98M. The effects of rotation should not 
be neglected in solar models. However, the internal angular velocity and 
total angular momentum of the rotating model are higher than
the seismically inferred values. A mechanism mainly taking effect in inner 
layers of models' radiative region is needed to solve the fast rotation problem.

\acknowledgments The author thanks the anonymous referee for helpful comments
that helped the author improve this work and Daniel Kister
for help in improving the English, and also acknowledges support from
the NSFC 11773005 and U1631236.

 \clearpage

 \clearpage

\clearpage

\clearpage

\clearpage

\clearpage

\begin{table*}
\begin{center}
\caption{Model parameters. \label{tab1}}
\begin{tabular}{p{1.4cm}p{0.9cm}p{1.1cm}lp{0.3cm}p{0.3cm}llllllp{0.2cm}l}
\hline\hline
 Model & $X_{0}$ & $Z_{0}$ & $\alpha$ &$\delta_{ov}$ & $f_{0}$ & $T_{c}$ & $\rho_{c}$ &$r_{cz}$ &$Y_{s}$ & $Z_{s}$& $(Z/X)_{s}$& $\Omega_{i}$ &$V_{e}$\\
       &         &         &          &              &    &  &   &$R_{\odot}$&        &    &  &   &    \\
\hline
 GN93M & 0.7061  & 0.02021  & 2.1130  &  0           & 1.0    & 1.576   & 154.41    &  0.715    &   0.2436 & 0.0182 & 0.0246 & 0 &0  \\
 GS98M & 0.7038  & 0.01950  & 2.1957  &  0           &   1.0  &  1.579   & 154.66   &  0.716    &   0.2461 & 0.0175 & 0.0238 & 0 &0     \\
 \textbf{GS98Ma} & 0.7010  & 0.02010 & 2.2130 &  0           &   1.0  &  1.583    & 155.05   &  0.715    &   0.2483 & 0.0181 & 0.0246 & 0 &0 \\
\hline
 GS98x1 & 0.7051 & 0.01954  & 2.2365  &  0           &   1.1  &  1.579   & 154.82   &  0.716    &   0.2422 & 0.0174 & 0.0235 & 0 &0  \\
 GS98x2 & \textbf{0.7038} & \textbf{0.01990}  & \textbf{2.3110}  &  \textbf{0} & \textbf{1.5} & \textbf{1.587} & \textbf{156.35} & \textbf{0.712} & \textbf{0.2335} & \textbf{0.0171} & \textbf{0.0228} & \textbf{0} & \textbf{0}  \\
 GS98x3 & \textbf{0.7014} & \textbf{0.02084}  & \textbf{2.4730}  &  \textbf{0} & \textbf{2.0} & \textbf{1.598} & \textbf{158.40} & \textbf{0.707} & \textbf{0.2240} & \textbf{0.0171} & \textbf{0.0225} & \textbf{0} & \textbf{0}  \\
\hline
 GS98r1& 0.7056 & 0.01925  & 2.1665  &  0.1 &   1.0  &  1.576   & 154.34   &  0.712 & 0.2548 & 0.0174 & 0.0238 & 9.4& 1.93 \\
 GS98r2& \textbf{0.7045} & \textbf{0.01958}  & \textbf{2.2392}  &  0.1 &   1.5  &  \textbf{1.584}   & \textbf{156.16}   &  \textbf{0.710} & \textbf{0.2466} & \textbf{0.0168} & \textbf{0.0229} & \textbf{10} & \textbf{2.01} \\
\hline
 AGSS1  & 0.7179 & 0.01588  &  2.1801  &  0   &   1.0  &  1.564   & 152.04   &  0.727    &  0.2351   & 0.0142 & 0.0189  & 0 & 0\\
 AGSSx2  & \textbf{0.7119} & \textbf{0.01686} &  \textbf{2.3654}  &  \textbf{0.1} & \textbf{1.5} & \textbf{1.578} & \textbf{154.33} & \textbf{0.713} & \textbf{0.2283} & \textbf{0.0144} & \textbf{0.0191} & \textbf{0} & \textbf{0}\\
 AGSSx3 & 0.7068 & 0.01772 & 2.4585  &   0    &  2.0   &  1.593   & 156.60   &  0.715    &  0.2204   & 0.0145 & 0.0189  & 0 & 0\\
 \textbf{AGSSx2a} & 0.7022& 0.01818  &  2.3577  &  0   &   1.5  &  1.592   & 155.51   &  0.716    &  0.2355   & 0.0155 & 0.0207  & 0 & 0\\ 
 \hline
 AGSSr2 & \textbf{0.7107} & \textbf{0.01700} & \textbf{2.3160} & \textbf{0.1} & \textbf{1.5} & \textbf{1.578}  & \textbf{154.42} & \textbf{0.715} & \textbf{0.2425} & \textbf{0.0145} & \textbf{0.0196} & \textbf{10} & \textbf{1.99} \\
 AGSSr3 & 0.7090  & 0.01740 & 2.3890  & 0.1 & 2.0 & 1.588 & 156.31  & 0.712 & 0.2354  & 0.0142 & 0.0189 & 10  & 2.02 \\
 \textbf{AGSSr2a} & 0.7050  & 0.01779 & 2.3098  & 0.1 & 1.5 & 1.586 & 155.13  & 0.713 &  0.2472 & 0.0152 & 0.0207 & 10 & 1.98 \\
 \hline
\end{tabular}
\tablenotetext{}{\textbf{Notes.} The central temperature $T_{c}$, central density $\rho_{c}$, initial angular velocity $\Omega_{i}$, 
and surface equatorial velocity $V_{e}$ are in unit of $10^{7}$ K, g cm$^{-3}$, $10^{-6}$ rad s$^{-1}$, and km s$^{-1}$, respectively.
}
\end{center}
\end{table*}

\begin{table*}
\begin{center}
\caption{Predicted solar neutrino fluxes from models.
\label{tab2}}
\begin{tabular}{ccccccccc}
\tableline\tableline
 model  &  $pp$ &  $pep$ &  $hep$ & $^{7}$Be & $^{8}$B & $^{13}$N & $^{15}$O & $^{17}$F  \\
 \tableline
 GN93M  & 5.95  & 1.40  &  9.51   & 5.03     & 5.10    & 5.77     & 5.05     & 5.59   \\
 GS98M &  5.95  & 1.40  &  9.46   & 5.13     & 5.25    & 5.50     & 4.87     & 5.64   \\
 \textbf{GS98Ma} & \textbf{5.93} & \textbf{1.40} & \textbf{9.40} & \textbf{5.23} & \textbf{5.47} & \textbf{5.85} & \textbf{5.21} & \textbf{6.04} \\
\tableline
 GS98x1 & 5.95  & 1.40  &  9.47   & 5.12     & 5.23    & 5.51     & 4.87     & 5.65    \\
 GS98x2 & \textbf{5.92}  & \textbf{1.39}  &  \textbf{9.36} & \textbf{5.31} & \textbf{5.67} & \textbf{6.10} & \textbf{5.45} & \textbf{6.34}    \\
 GS98x3 & \textbf{5.87}  & \textbf{1.38}  &  \textbf{9.21} & \textbf{5.59} & \textbf{6.35} & \textbf{7.17} & \textbf{6.49} & \textbf{7.59}   \\
\tableline
 GS98r1 & 5.95  & 1.40 &  9.46    & 5.00     & 5.00    & 5.26     & 4.63     & 5.35 \\
 GS98r2 & \textbf{5.92}  & \textbf{1.40} &  \textbf{9.34}    & \textbf{5.21}     & \textbf{5.47}    & \textbf{5.88} & \textbf{5.23} & \textbf{6.08}  \\
\tableline
 AGSS1  & 6.01   &  1.43 &  9.77   & 4.70     & 4.38    & 3.88     & 3.36     & 3.86     \\
 AGSSx2 & \textbf{5.96}   & \textbf{1.41}  &  \textbf{9.54}   & \textbf{5.07}     & \textbf{5.16}    & \textbf{4.78}     & \textbf{4.22} & \textbf{4.89} \\
 AGSSx3 & 5.91   & 1.39  &  9.33   & 5.45     & 6.03    & 5.82     & 5.23     & 6.11     \\
 \textbf{AGSSx2a} & 5.91   & 1.39  &  9.31   & 5.43     & 5.99    & 5.80     & 5.22     & 6.08     \\
 \tableline
 AGSSr2 & \textbf{5.94}  & \textbf{1.40}  &  \textbf{9.47}   & \textbf{5.04}     & \textbf{5.11}    & \textbf{4.81}     & \textbf{4.25} & \textbf{4.92} \\
 AGSSr3 &  5.90  & 1.39  &  9.34   & 5.28     & 5.66    & 5.47     & 4.89   & 5.70  \\
 \textbf{AGSSr2a} &  5.92  & 1.39  &  9.34   & 5.25     & 5.60    & 5.42     & 4.84   & 5.62  \\
 \tableline
 BP04 &  5.94  & 1.40   &  7.88   & 4.86     & 5.79    & 5.71     & 5.03     & 5.91     \\
 SSeM &  5.92  & 1.39   &  ....   & 4.85     & 4.98    & 5.77     & 4.97     & 3.08    \\
 Measured & 6.06$^{+0.02 a}_{-0.06}$ & 1.6$\pm$0.3$^{b}$ &...& 4.84$\pm$0.24$^{a}$& 5.21$\pm$0.27$\pm$0.38$^{c}$ &...& ...&...\\
\tableline
\end{tabular}
\tablenotetext{}{\textbf{Notes.} The table shows the predicted fluxes, in units of $10^{10}(pp)$,
$10^{9}(^{7}\mathrm{Be})$, $10^{8}(pep,^{13}\mathrm{N}, ^{15}\mathrm{O})$,
$10^{6}(^{8}\mathrm{B}, ^{17}\mathrm{F})$, and $10^{3}(hep)$
$\mathrm{cm}^{2} \mathrm{s}^{-1}$. The BP04 is the best model of \cite{bah04a} and
has the GS98 mixtures. The SSeM is a standard model with high metal abundances
that reproduces the seismic sound speed \citep{cou03, tur11a}.}
\tablenotetext{a}{\cite{bel11}.}
\tablenotetext{b}{\cite{bel12}.}
\tablenotetext{c}{\cite{ahm04}.}
\end{center}
\end{table*}


\begin{thebibliography}{}

\bibitem[Adelberger et al.(1998)]{ade98} Adelberger, E. C., Austin, S. M., Bahcall, J. N. et al. 1998, Rev. Mod. Phys., 70, 1265
\bibitem[Ahmed et al.(2004)]{ahm04} Ahmed, S. N., Anthony, A. E., Beier, E. W. et al. 2004, PhRvL, 92, 1301
\bibitem[Anders \& Grevesse(1989)]{and89} Anders, E., \& Grevesse, N. 1989, GeCoA, 53, 197
\bibitem[Asplund et al.(2004)]{asp04} Asplund, M., Grevesesse N., Sauval, A. J., Allende Prieto, C., \& Kiselman, D. 2004, A\&A, 417, 751
\bibitem[Asplund et al.(2009)]{asp09} Asplund, M., Grevesse, N., Sauval, A.,
\& Scott, P. 2009, ARA\&A, 47, 481 (AGSS09)

\bibitem[Bahcall et al.(2001)]{bah01} Bahcall, J. N., Pinsonneault, M. H.,
\& Basu, S, 2001, ApJ, 555, 990
\bibitem[Bahcall \& Pinsonneault(1992)]{bah92} Bahcall, J. N., \& Pinsonneault, M. H.
1992, Rev. Mod. Phys., 64, 885
\bibitem[Bahcall et al.(1995)]{bah95} Bahcall J. N., Pinsonneault M. H.,
\& Wasserburg G. J. 1995, Rev. Mod. Phys., 67, 781
\bibitem[Bahcall \& Pinsonneault(2004)]{bah04a} Bahcall, J. N., \& Pinsonneault, M. H.
2004, Phys. Rev. Lett., 92, 121301
\bibitem[Bahcall et al.(2004)]{bah04b} Bahcall, J. N., Serenelli, A. M., \&
Pinsonneault, M. H. 2004, ApJ, 614,464
\bibitem[Bahcall et al.(2005)]{bah05} Bahcall, J. N., Basu, S.,
Pinsonneault, M. H., \& Serenelli, A. M. 2005, ApJ, 618, 1049
\bibitem[Basu \& Antia(1997)]{bas97} Basu, S., \& Antia, H. M.
1997, MNRAS, 287, 189
\bibitem[Basu et al.(2009)]{bas09} Basu, S., Chaplin, W. J., Elsworth, Y.,
New, R., Serenelli, A. M. 2009, ApJ, 699, 1403
\bibitem[Basu \& Antia(2004)]{bas04} Basu, S., \& Antia, H. M.
2004, ApJ, 606, L85
\bibitem[Bellini et al.(2011)]{bel11} Bellini, G. Benziger, J. Bick, D. et al. 2011, PhRvL, 107, 141302
\bibitem[Bellini et al.(2012)]{bel12} Bellini, G. Benziger, J. Bick, D. et al. 2012, PhRvL, 108, 51302
\bibitem[Bi et al.(2011)]{bi11} Bi, S. L., Li, T. D., Li, L. H., Yang, W. M. 2011, ApJL, 731, 42
\bibitem[Brott et al.(2011)]{bro11} Brott, I., Evans, C. J., Hunter, I., et al. 2011, A\&A, 530, A116
\bibitem[Buldgen et al.(2017)]{buld17} Buldgen, G.,Salmon, S. J. A. J., Noels, A., Scuflaire, R.,
Dupret, M. A., Reese, D. R. 2017, MNRAS, 472, 751

\bibitem[Caffau et al.(2010)]{caf10} Caffau, E., Ludwig, H.-G.,
Bonifacio, P., Faraggiana, R., Steffen, M., Freytag, B., Kamp, I., Ayres, T. R. 2010, A\&A, 514, A92
\bibitem[Castro et al.(2007)]{cas07} Castro, M., Vauclair, S., Richard, O. 2007, A\&A, 463, 755
\bibitem[Chaboyer et al(1995)]{cha95} Chaboyer, B., Demarqure, P., Pinsonneault, M. H. 1995, 441, 865
\bibitem[Chaboyer \& Zahn(1992)]{cha92} Chaboyer, B., \& Zahn, J. P. 1992, A\&A, 253, 173
\bibitem[Chaplin et al.(1999a)]{cha99a} Chaplin, W. J., Christensen-Dalsgaard, J., Elsworth, Y., et al. 1999a, MNRAS, 308, 405
\bibitem[Chaplin et al.(1996)]{cha96} Chaplin, W. J., Elsworth, Y., Howe, R., et al. 1996, Sol. Phys., 168, 1
\bibitem[Chaplin et al.(1999)]{cha99b}
Chaplin, W. J., Elsworth, Y., Isaak, G. R., Miller, B. A., \& New, R. 1999b, MNRAS, 308, 424

\bibitem[Charbonnel \& Talon(2005)]{cha05} Charbonnel, C., \& Talon, S. 2005, Science, 309, 2189
\bibitem[Christensen-Dalsgaard et al.(1991)]{chr91}
Christensen-Dalsgaard, J., Gough, D. O., \& Thompson, M. J. 1991, ApJ, 378, 413
\bibitem[Couvidat et al.(2003)]{cou03} Couvidat, S., Turck-Chi$\grave{e}$ze, S., \& Kosovichev, A. G. 2003, ApJ, 599, 1434
\bibitem[Denissenkov et al.(2008)]{den08} Denissenkov, P. A., Pinsonneault, M., MacGregor, K. B. 2008, ApJ, 684, 757
\bibitem[Endal \& Sofia(1978)]{end78} Endal, A. S., \& Sofia, S. 1978, ApJ, 220, 279
\bibitem[Ferguson et al.(2005)]{fer05} Ferguson, J. W., Alexander, D. R., Allard, F. et al. 2005, ApJ, 623, 585
\bibitem[Garc\'{\i}a et al.(2011)]{gar11}
Garc\'{\i}a, R. A., Salabert, D., \& Ballot, J. et al., 2011, JPhCS, 271, 012049
\bibitem[Garc\'{\i}a et al.(2007)]{gar07} Garc\'{\i}a, R. A., Turck-Chi$\grave{e}$ze, S., Jim$\acute{e}$nez-Reyes,
S. J., Ballot, J., Pall$\acute{e}$, P. L., Eff-Darwich, A., Mathur, S., Provost, J. 2007, Sci., 316, 1591

\bibitem[Grevesse \& Noels(1993)]{gre93} Grevesse, N., Noels, A. 1993. In Origin and Evolution of the Elements,
ed. N Prantzos, E Vangioni-Flam, MCass\'{e}, pp. $15-25$. Cambridge, UK: Cambridge Univ. Press. 561 pp. (GN93)
\bibitem[Grevesse \& Sauval(1998)]{gre98} Grevesse, N., \& Sauval, A. J. 1998,
in Solar Composition and Its Evolution, ed. C. Fr$\ddot{o}$hlich, et al. (Dordrecht: Kluwer), 161 (GS98)

\bibitem[Gruzinov \& Bahcall(1998)]{gru98} Gruzinov, A. V., \& Bahcall, J. N. 1998, ApJ, 504, 996
\bibitem[Guenther(1994)]{gue94} Guenther D. B., 1994, ApJ, 422, 400
\bibitem[Guzik et al.(2005)]{guz05} Guzik, J. A., Watson, L. S. \&
Cox, A. N. 2005, ApJ, 627, 1049
\bibitem[Guzik et al.(2010)]{guz10} Guzik, J. A., \& Mussack, K. 2010,
ApJ, 713, 1108

\bibitem[Iglesias \& Rogers(1996)]{igl96} Iglesias, C., Rogers, F. J. 1996, ApJ, 464, 943
\bibitem[Kawaler(1988)]{kaw88} Kawaler, S. D. 1988, ApJ, 333, 236
\bibitem[Komm et al.(2003)]{kom03} Komm, R., Howe, R., Durney, B. R., \& Hill, F. 2003, ApJ, 586, 650
\bibitem[Le Pennec et al.(2015)]{lep15} Le Pennec, M., Turck-Chi$\grave{e}$ze, S.,
Salmon, S., Blancard, C., Coss$\acute{e}$, P., Faussurier, G., Mondet, G. 2015, ApJL, 813, L42
\bibitem[Lodders(2003)]{lod03} Lodders, K. 2003, ApJ, 591, 1220
\bibitem[Lodders et al.(2009)]{lod09} Lodders, K, Palme, H, Gail, H-P. 2009, LanB, 4, 44
\bibitem[Lopes \& Turck-Chi$\grave{e}$ze(2013)]{lop13} Lopes, I., \& Turck-Chi$\grave{e}$ze, S. 2013, ApJ, 765, 14
\bibitem[Lopes \& Turck-Chi$\grave{e}$ze(2014)]{lop14} Lopes, I., \& Turck-Chi$\grave{e}$ze, S. 2014, ApJL, 792, L35
\bibitem[Maeder \& Zahn(1998)]{mae98} Maeder, A., \& Zahn, J.-P. 1998, A\&A, 334, 1000
\bibitem[Marcucci et al.(2000)]{mar00} Marcucci, L. E., Schiavilla, R., Viviani, M.,
Kievski, A., \& Rosati, S. 2000, Phys. Rev. Lett., 84, 5959
\bibitem[Montalb$\acute{a}$n et al.(2004)]{mon04} Montalb$\acute{a}$n, J., Miglio, A.,
Noels, A., Grevesse, N., Di Mauro, M. P. 2004, in Helio- and
Asteroseismology: Towards a Golden Future, Proc. of the SOHO 14 /
GONG 2004 Workshop, ed. D. Danesy (ESA SP-559; Noordwijk: ESA), 574
\bibitem[Montalb$\acute{a}$n et al.(2006)]{mon06} Montalban, J., Miglio, A.,
Theado, S., Noels, A., Grevesse, N. 2006, Commun. Asteroseismol., 147, 80

\bibitem[Pinsonneault et al.(1989)]{pin89} Pinsonneault, M. H., Kawaler, S. D., Sofia, S., \&
Demarqure, P., 1989, ApJ, 338, 424
\bibitem[Proffitt \& Michaud(1991)]{pro91} Proffitt, C. R., \& Michaud, G. 1991, ApJ, 380, 238
\bibitem[Rogers \& Nayfonov(2002)]{rog02} Rogers, F., \& Nayfonov, A. 2002, ApJ, 576, 1064
\bibitem[Schou et al.(1998)]{sch98} Schou, J., Antia, H. M., Basu, S. et al. 1998, ApJ, 505, 390
\bibitem[Serenelli \& Basu(2010)]{ser10} Serenelli, A., \& Basu, S. 2010, ApJ, 719, 865
\bibitem[Serenelli et al.(2009)]{ser09} Serenelli, A., Basu, S., Ferguson, J., \& Asplund, M. 2009, ApJL, 705, 123
\bibitem[Serenelli et al.(2011)]{ser11}
Serenelli, A., Haxton, W. C., Pe$\tilde{n}$a-Garay, C. 2011, ApJ, 743, 24
\bibitem[Serenelli et al.(2016)]{ser16} Serenelli, A. M., Scott, P., Villante, F. L., et al. 2016, MNRAS, 463, 2

\bibitem[Thoul, Bahcall \& Loeb(1994)]{tho94} Thoul, A. A., Bahcall, J. N.,
Loeb, A. 1994, ApJ, 421, 828
\bibitem[Turck-Chi$\grave{e}$ze \& Couvidat(2011)]{tur11a} Turck-Chi$\grave{e}$ze, S., \& Couvidat, S. 2011, RPPh, 74, 6901
\bibitem[Turck-Chi$\grave{e}$ze et al.(2004)]{tur04} Turck-Chi$\grave{e}$ze, S.,
Couvidat, S., Piau, L., Ferguson, J., Lambert, P., Ballot, J., Garc$\acute{i}$a, R. A.,
Nghiem, P. 2004, Phys. Rev. Lett., 93, 211102
\bibitem[Turck-Chi$\grave{e}$ze \& Lopes(2012)]{tur12} Turck-Chi$\grave{e}$ze, S., \& Lopes, I. 2012, RAA, 12, 1107
\bibitem[Turck-Chi$\grave{e}$ze et al.(2010)]{tur10} Turck-Chi$\grave{e}$ze, S., Palacios, A., Marques, J. P.,
Nghiem, P. A. P. 2010, ApJ, 715, 1539
\bibitem[Turck-Chi$\grave{e}$ze et al.(2011)]{tur11} Turck-Chi$\grave{e}$ze, S., Piau, L., \& Couvidat, S.
2011, ApJL, 731, L29

\bibitem[Vagnozzi et al.(2017)]{vag17} Vagnozzi, S., Freese, K., Zurbuchen, T. H. 2017, ApJ, 839, 55
\bibitem[von Steiger \& Zurbuchen(2016)]{von16} von Steiger, R., \& Zurbuchen, T. H. 2016, ApJ, 816, 13
\bibitem[Vorontsov et al.(2014)]{voro14} Vorontsov, S. V., Baturin, V. A., Ayukov, S. V., Gryaznov, V. K. 2014, MNRAS, 441, 3296

\bibitem[Yang \& Bi(2006)]{yang06} Yang, W. M., \& Bi, S. L. 2006, \aap, 449, 1161
\bibitem[Yang \& Bi(2007)]{yang07} Yang, W. M., \& Bi, S. L. 2007, ApJ, 658, L67
\bibitem[Yang et al.(2013)]{yang13} Yang, W., Bi, S., Meng, X., \& Liu, Z 2013, ApJ, 776, 112
\bibitem[Yang(2016)]{yang16} Yang, W. 2016, ApJ, 821, 108
\bibitem[Zahn(1993)]{zahn93} Zahn, J. P. 1993, in Astrophysical Fluid Dynamics, Les Houches XLVII, ed.
J.-P. Zahn \& Zinn-Justin (New York: Elsevier), 561
\end{thebibliography}
\end{document}